\documentclass[final,1p,times]{elsarticle} 
\usepackage{graphicx} 
\usepackage{amssymb} 
\usepackage{amsthm} 
\usepackage{lineno} 

\journal{Nuclear Physics A} 
\begin{document} 

\begin{frontmatter} 


\title{Jet evolution in Yang-Mills-Wong simulations}

\author{Bj\"orn Schenke$^{a}$}

\address[a]{Department of Physics, McGill University, Montreal, Quebec, H3A 2T8, Canada}

\begin{abstract} 
We present results for collisional energy loss and momentum broadening of high momentum partons in a hot and dense non-Abelian plasma 
obtained by solving the coupled system of Yang-Mills-Wong equations on a lattice in real time. Including hard elastic collisions among 
the particles we obtain cutoff independent results for the collisional energy loss $dE/dx$ and the transport coefficient $\hat{q}$. 
The latter is found to receive a sizable contribution from a power-law tail in the transverse momentum distribution of high-momentum 
partons.
We further argue that the effect of instabilities on jet broadening should be accessible by experiment
when employing jet cones with elliptical bases or studying correlations within the cone
in full jet reconstruction methods.
\end{abstract} 

\end{frontmatter} 



\section{Introduction}
High transverse momentum jets produced in heavy-ion collisions represent a valuable
tool for studies of the properties of the hot parton plasma produced in the central
rapidity region \cite{Jacobs:2005pk}. 
We study the collisionsal energy loss and momentum broadening of high momentum partons 
employing a numerical simulation of the Boltzmann-Vlasov equation, which is coupled to the Yang-Mills equation for the soft
gluon degrees of freedom. Soft momentum exchanges between particles are mediated by
the fields, while hard momentum exchanges are described by a collision term including
binary elastic collisions. This way, we are able to provide an estimate of the coupling of
jets to a hot plasma which is independent of infrared cutoffs.

We also simulate plasmas with local momentum-space anisotropies,
which occur in heavy-ion collisions due to the longitudinal expansion
during the very early stages of the plasma evolution.
These plasmas develop Chromo-Weibel instabilities~\cite{Romatschke:2003ms},
which lead to the formation of long-wavelength
chromo-fields with $E_z>B_z$ and $B_\perp >E_\perp$.
The strong fields affect the propagation of hard partons, leading to
an asymmetry of the jet shape in rapidity $\Delta\eta$ and azimuthal angle $\Delta\phi$,
which should be accessible using new jet reconstruction measurements \cite{jets}.

\section{Boltzmann-Vlasov equation for non-Abelian gauge theories}
We solve the classical transport equation for hard gluons with SU(2)
color charge 
including hard binary collisions
\begin{eqnarray}
    p^\mu \left[\partial_\mu + g q^a F_{\mu\nu}^a \partial^\nu_{p} 
    + g f^{abc} A_\mu^b(x) q^c \partial_{q^a} \right]f={\cal C}\,,
\end{eqnarray}
where $f=f(x,p,q)$ denotes the single-particle phase space distribution.
It is coupled self-consistently to the Yang-Mills equation for the
soft gluon fields
 \begin{equation}
  D_\mu F^{\mu\nu} = j^\nu 
  = g \int \frac{d^3p}{(2\pi)^3}\,dq\,q\,v^\nu\,f(x,p,q)\,,
 \end{equation}
with $v^\mu=(1,\mathbf{p}/p)$. 
The collision term contains all binary collisions, descibed 
by the leading-order $gg\rightarrow gg$ tree-level diagrams.

We replace the distribution $f(x,p,q)$ by a large number of test
particles, which leads to Wong's equations~\cite{Wong:1970fu}
\begin{eqnarray}
    \dot{\mathbf{x}}_i(t)&=&\mathbf{v}_i(t)\,,\nonumber\\
    \dot{\mathbf{p}}_i(t)&=&g q^a_i(t)\left(\mathbf{E}^a(t)+\mathbf{v}_i(t)\times\mathbf{B}^a(t)\right)
    \,,\nonumber \\
    \dot{q}_i(t)&=&-igv_i^\mu(t)[A_\mu(t),q_i(t)]\,,\label{wong}
\end{eqnarray}
for the $i$-th test particle, whose coordinates are $\mathbf{x}_i(t)$,
$\mathbf{p}_i(t)$, and $q^a_i(t)$, the color charge. The time evolution of the
Yang-Mills field is determined by the standard Hamiltonian method
\cite{Ambjorn:1990pu} in $A^0=0$ gauge. See~\cite{HuBM,Dumitru:2005gp,Dumitru:2006pz,Dumitru:2007rp,Schenke:2008gg}
for more details.
The collision term is incorporated using the stochastic method
\cite{Danielewicz:1991dh}. 
The total cross section is given by
$
    \sigma_{2\to2}=\int_{k^{*2}}^{{s}/{2}}\frac{d\sigma}{dq^2}dq^2\,,
$
where we have introduced a lower cutoff $k^*$.
To avoid double-counting, this cutoff
should be on the order of the hardest field mode that can be
represented on the given lattice, $k^*\simeq\pi/a$, with the lattice spacing $a$.
This way soft momentum exchanges are mediated by the fields while hard momentum exchanges are
described by the collision term.
For a more detailed discussion see \cite{Schenke:2008gg}.

\section{Jet broadening in an isotropic plasma}
We first consider heat-baths of particles with different densities $n_g$ and temperatures $T$.
We simulate an undersaturated gluon plasma due to numerical restrictions (see \cite{Schenke:2008gg})
and later extrapolate to thermal systems with physical densities and temperatures.
For a given lattice (or $k^*$) we take the initial
energy density of the thermalized fields to be $\int d^3k/(2\pi)^3 \,
k \hat{f}_{\rm Bose}(k)\Theta(k^*-k)$, where $\hat{f}_{\rm Bose}(k)=n_g/(2T^3 \zeta(3))/(e^{k/T}-1)$ is a Bose
distribution normalized to the assumed particle density $n_g$, and $\zeta$ is the Riemann zeta function.
The initial field spectrum is fixed to Coulomb gauge and $A_i\sim 1/k$.

\begin{figure}[t]
  \hfill
  \begin{minipage}[ht]{.48\textwidth}
  \begin{center}
    \includegraphics[width=6.5cm]{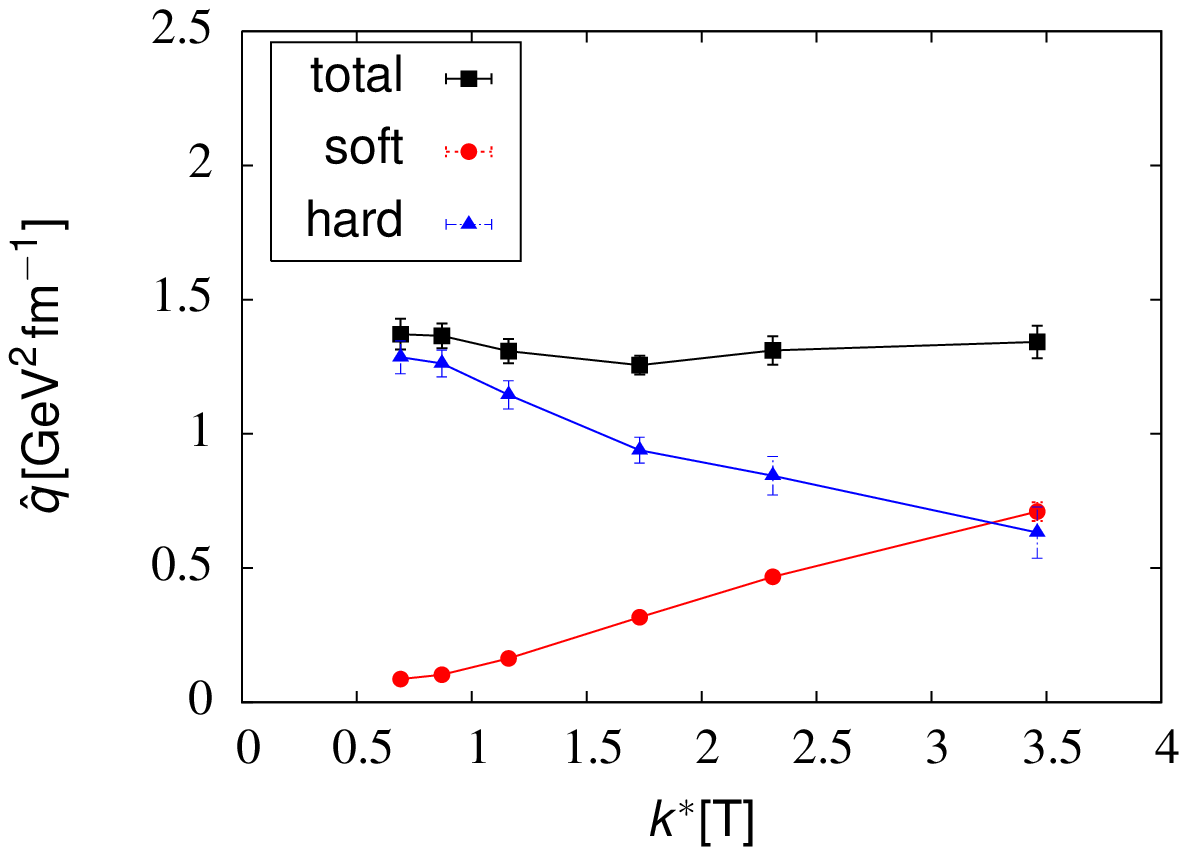}
  \end{center}
  \end{minipage}
  \hfill
  \begin{minipage}[ht]{.48\textwidth}
  \begin{center}
    \includegraphics[width=6.7cm]{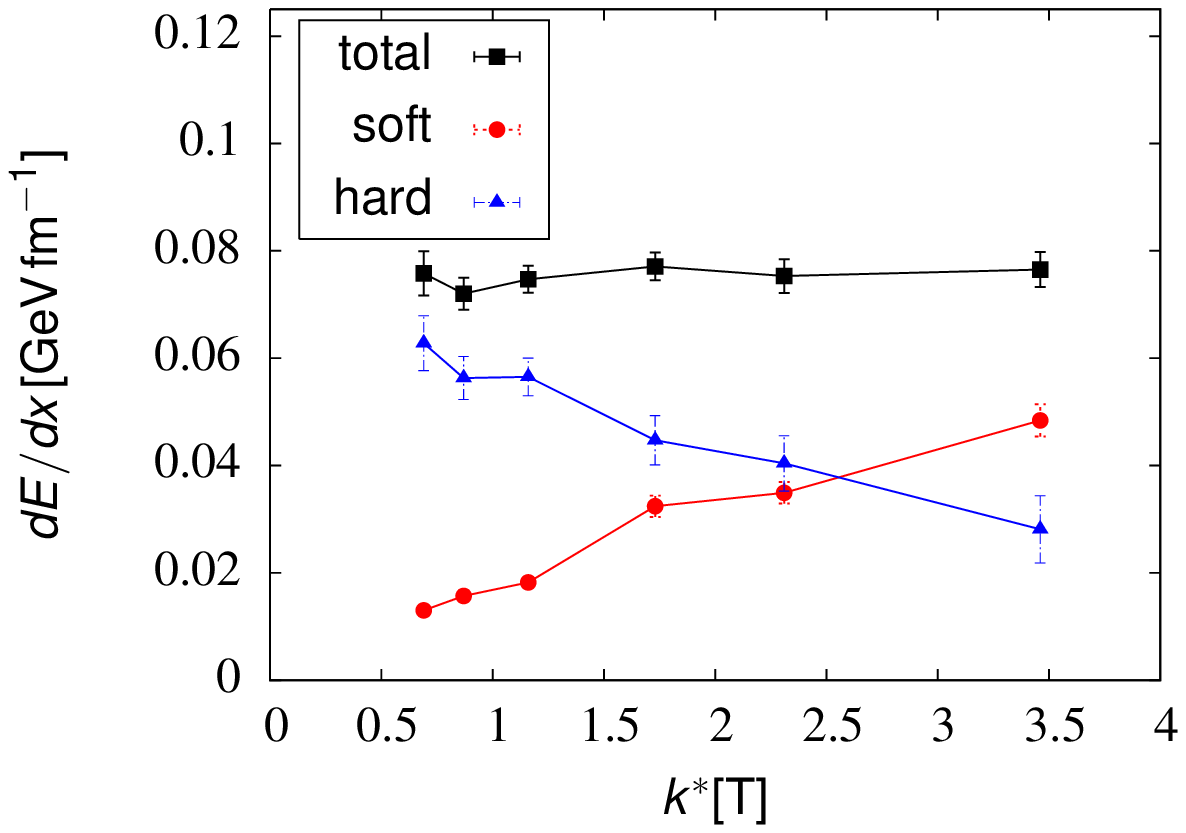}
   \end{center}
  \end{minipage}
  \caption{(In-)dependence of the transport coefficient $\hat{q}$ (left) and the elastic energy loss (right)
      on the separation scale $k^*$. $T=4$ GeV, $g=2$,
      $n=5\,\mathrm{fm}^{-3}$, $E=48\,T$. \label{fig:qhatk}}
    \hfill
\end{figure}

We study colorless bunches of test particles that allow us to restrict to collisional 
energy loss and momentum broadening due to elastic collisions only. Both
\begin{equation}\label{eq:qhat}
 	\hat{q}=\frac{1}{\lambda \sigma} \int
        d^2p_\perp\,p_\perp^2\frac{d\sigma}{d p_\perp^2}=\frac{\langle p_\perp^2\rangle(t)}{t}\,,
\end{equation}
and the differential energy loss $dE/dx$ are independent of the
separation scale $k^*$. $\lambda$ is the mean free path and $p_\perp$ denotes
the momentum transverse to the initial jet momentum.
Fig.~\ref{fig:qhatk} depicts the contributions to $\hat{q}$ and $dE/dx$ due to
soft and hard collisions separately, as well as the total.

Studying the dependencies of $\hat{q}$ and $dE/dx$ on the density, temperature, and hard parton energy
reproduces results from perturbative QCD (see \cite{Schenke:2008gg} for a detailed analysis).
Using these dependencies we can extrapolate $\hat{q}$ and $dE/dx$
to temperatures around $400$~MeV. Adjusting the color factors as appropriate
for SU(3) and extrapolating to the thermal density of gluons we
find $\hat{q}\approx 3.6\pm 0.3\,\mathrm{GeV}^{2}\,\mathrm{fm}^{-1}$ and
$dE/dx\approx 1.6\pm 0.4$ GeV fm$^{-1}$ ($E=19.2\,\mathrm{GeV}$).

Next, we present the full $p_\perp^2$-distribution of the
high-momentum partons traversing the hot medium in order to assess the
relative contributions from various processes to its first moment
$\hat{q}$. We find that over time the initial $\delta$-function
broadens to a Gaussian distribution with a power-law tail. 
\begin{figure}[t]
  \hfill
  \begin{minipage}[ht]{.48\textwidth}
  \begin{center}
    \includegraphics[width=6.5cm]{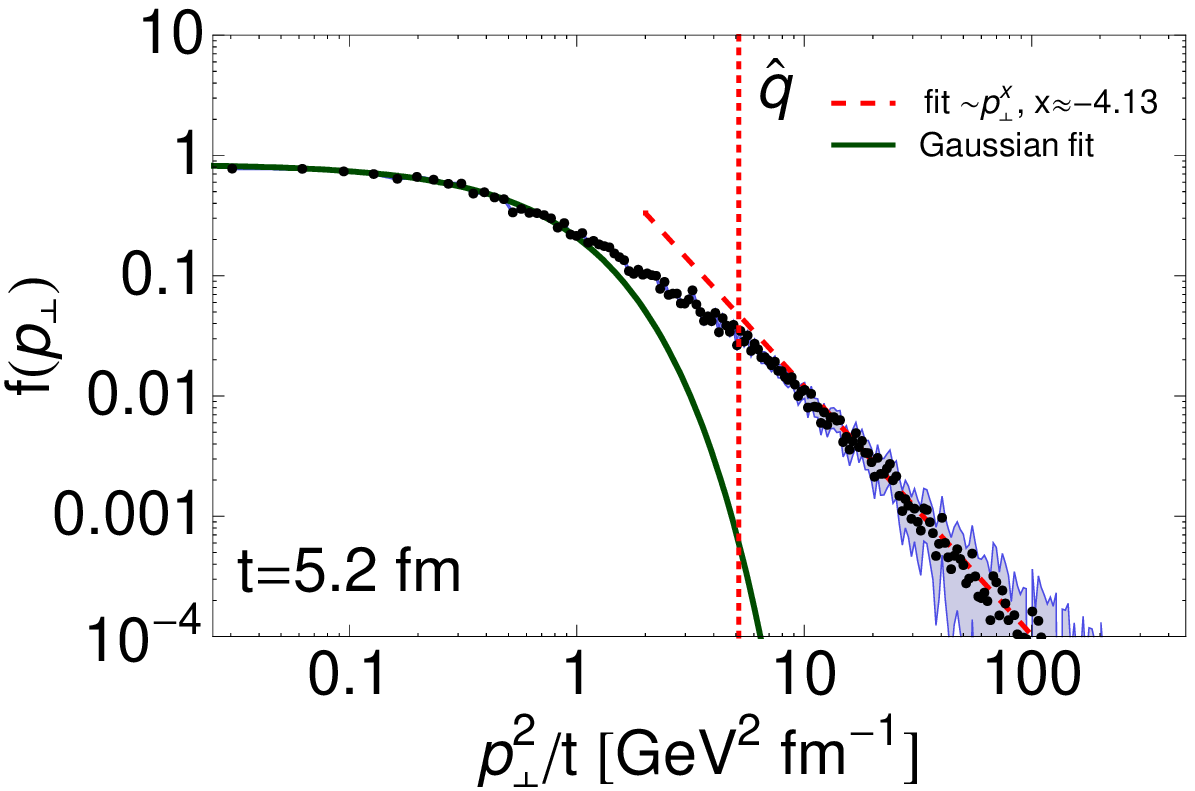}
    \caption{$p_\perp^2$-distribution of the high-momentum ($E/T=48$)
      partons after $t\approx 5.2\,\mathrm{fm}$ for
      $T=4\,\mathrm{GeV}$ and $n=20\,\mathrm{fm}^{-3}$
      ($\hat{q}\approx 5.16\,\mathrm{GeV}^2\mathrm{fm}^{-1}$). \label{fig:ptdistlogbin-n20-5.2fm} }
  \end{center}
  \end{minipage}
  \hfill\hspace{0.1cm}
  \begin{minipage}[ht]{.48\textwidth}
  \begin{center}
    \includegraphics[width=6.5cm]{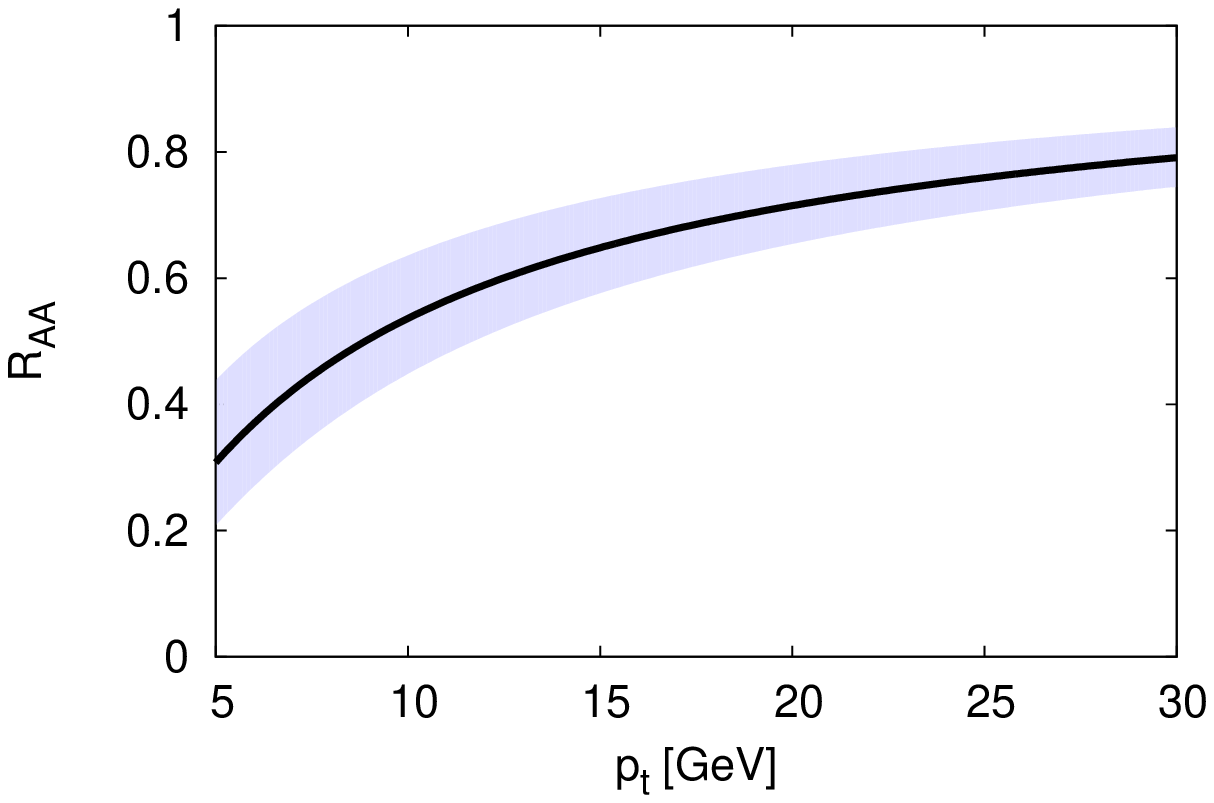}
    \caption{Nuclear modification factor
      $R_{AA}(p_\perp)$ of jets due to elastic energy loss in a
      classical Yang-Mills field produced in the early stage of a
      relativistic heavy-ion collision at RHIC. The band indicates the
      uncertainty due to the extrapolation of $dE/dx$ to
      physical temperatures. \label{fig:R_AA}}
  \end{center}
  \end{minipage}
    \hfill
\end{figure}
Fig.~\ref{fig:ptdistlogbin-n20-5.2fm} shows the distribution of the
high-momentum test particles after $t\approx 5.2\, \mathrm{fm}$ in a
double-logarithmic plot versus $p_\perp^2/t$. 
The low-$p_\perp$ part follows a Gaussian distribution in $p_\perp$.
The power-law tail at large $p_\perp$ behaves approximately as $p_\perp^{-4}$.  This
is expected for particles experiencing only few scatterings since in
the high-energy limit the differential cross section
$d\sigma/dp_\perp^2 \sim p_\perp^{-4}$. We also indicate the value of
$\hat{q}$ to show that the power-law tail contributes significantly to
this transport coefficient.

We also provide an estimate for the nuclear modification
factor $R_{AA}$ due to elastic energy loss in a
classical Yang-Mills field.
This is of relevance for collisions
of heavy nuclei at high energies: the large number of gluons produced
in the central rapidity region can be described as a classical field
for a short time~(see e.g. \cite{Krasnitz:1999wc})
until the field modes decohere and thermalize~\cite{Dumitru:2006pz}. 
From the simulations presented above, 
we obtained the elastic energy loss $dE/dx$.
For large $k^*$ (on the order of the ``saturation
momentum'' $Q_s$) most of the energy density is due to the classical
field. In a thermal system at $T=400$~MeV the energy density is about 17~GeV/fm$^3$,
which is an appropriate average over the first 1~fm/c of a central
Au+Au collision at RHIC energy~\cite{Krasnitz:1999wc}.
An estimate for the nuclear modification factor $R_{AA}$ at the parton level
(neglecting hadronization) can be written as~\cite{Wicks:2005gt,Schenke:2008gg}
\begin{equation} R_{AA}(p_\perp) = \frac{dN_f/d^2p_\perp dy}{dN_i/d^2p_\perp dy}
=\left( 1-\epsilon(p_\perp) \right)^n ~.  
\end{equation} 
$\epsilon$ denotes the fractional energy loss up to $\tau=1\,{\rm fm}$ 
and $dN_i/d^2p_\perp dy \sim 1/p_\perp^{n+2}$ is the initial $p_\perp$
distribution of jets, where $n\approx 4$.
Fig.~\ref{fig:R_AA} shows the result for $R_{AA}$. The experimentally
observed flat $R_{AA}\approx0.2$ can not be fully accounted for by
early-stage elastic energy loss in the classical field
background, but our result shows that this contribution is
significant and can not be neglected.

\section{Jet broadening in an unstable plasma}
As studied in detail in \cite{Dumitru:2007rp}, instability growth in unstable plasmas
leads to a direction dependent $\hat{q}$. 
We describe the broadening of the hard (now colored) test particles
in the transverse $\perp$ and longitudinal $z$ directions via the variances
$
	\hat{q}_\perp:=\frac{d}{dt}\langle(\Delta p_\perp)^2\rangle\,, \hat{q}_z:=\frac{d}{dt}\langle(\Delta p_z)^2\rangle\,.
$
The ratio $\hat{q}_z/\hat{q}_\perp$ can be roughly associated with the
ratio of jet correlation widths in azimuth and rapidity:
$\sqrt{{\hat{q}_z}/{\hat{q}_\perp}} \approx {\langle\Delta\eta\rangle} /
{\langle\Delta\phi\rangle}$.
The numerical simulations of unstable plasmas obtain \cite{Dumitru:2007rp}
\begin{equation}
  {\langle\Delta\eta\rangle} / {\langle\Delta\phi\rangle}\approx 1.5\,.
\end{equation} 
We propose to look for this effect by using the greatly improved experimental methods for full jet reconstruction 
\cite{jets,Bruna:2009kw,Salur:2009rg}.
In particular, one should use cones with an elliptical base and determine its
major axis in the $\phi$-$\eta$ plane by maximizing the energy contained in the cone.
If the system is isotropic, the angle should be uniformly distributed while the effect of instabilities
should lead to a preferred orientation of the major axis in the $\eta$ direction. 
Another, possibly more practical way of experimentally determining such an anisotropy of the jet
is to keep the circular base and study correlations among the particles within the cone.

\section*{Acknowledgments} 
I thank Adrian\ Dumitru, Carsten\ Greiner, Yasushi\ Nara, and Michael\ Strickland for their collaboration
and Charles Gale and Sangyong Jeon for helpful discussions. I gratefully acknowledge a Richard H.~Tomlinson
Fellowship awarded by McGill University as well as support from the
Natural Sciences and Engineering Research Council of Canada.  
\end{document}